\pdfoutput=1 
\documentclass[screen,dvipsnames,nonacm,acmlarge]{acmart}
\usepackage[utf8]{inputenc}

\copyrightyear{2021}
\setcopyright{none}
\acmYear{}
\acmConference{}{}{}
\acmPrice{}
\acmDOI{}
\acmISBN{}

\usepackage{physics}
\usepackage{amsthm}
\usepackage{mathtools}  
\usepackage{csquotes}

\newcommand{\B}{B}

\newcommand{\bl}{b}

\DeclareMathOperator{\arctantwo}{arctan2}
\DeclareMathOperator{\sinhc}{sinch}
\DeclareMathOperator{\sinc}{sinc}

\DeclareMathOperator{\Log}{Log}
\DeclareMathOperator{\sign}{sign}
\DeclareMathOperator{\si}{\mathbf s}
\DeclareMathOperator{\co}{\mathbf c}
\DeclareMathOperator{\ta}{\mathbf t}

\newcommand{\sico}[3][{}]{\si^{#1}(#2)\co^{#1}(#3)}

\newcommand{\GR}[2][{}]{\ensuremath{\mathbb{R}^{{#1}}_{#2}}}
\newcommand{\ER}[1]{\ensuremath{\mathbb{R}^{{#1}}}}
\newcommand{\e}{\mathbf {e}_}            

\newcommand{\Spin}[1]{\ensuremath{\text{Spin}({#1})}}
\newcommand{\Pin}[1]{\ensuremath{\text{Pin}({#1})}}

\newcommand{\Eu}[1]{\ensuremath{\text{E}({#1})}}

\newcommand{\SO}[1]{\ensuremath{\text{SO}({#1})}}
\newcommand{\spin}[1]{\ensuremath{\mathfrak{spin}({#1})}}

\DeclareMathOperator\Ln{Ln}

\DeclarePairedDelimiter\floor{\lfloor}{\rfloor}

\DeclareMathOperator{\grade}{grade}

\usepackage{pdflscape} 
\usepackage{subcaption}
\usepackage[font=small,labelfont=bf]{caption}
\usepackage{colortbl}
\usepackage{accents}
\usepackage{eqnarray}
\usepackage{cleveref}
\usepackage{epigraph}
\usepackage{awesomebox}
\usepackage{cancel}
\usepackage{caption}
\usepackage[frozencache=true,cachedir=minted-cache]{minted} 
\setminted{fontsize=\footnotesize,autogobble}
\setcitestyle{unsrtnat}
\usepackage{footmisc}

\usepackage[h]{esvect}

\usepackage{pdfpages}

\usepackage{calc}  
\usepackage{enumitem}

\crefname{algocf}{algorithm}{algorithms}
\Crefname{algocf}{Algorithm}{Algorithms}
\crefname{figure}{fig.}{figs.}
\Crefname{figure}{Figure}{Figures}

\newenvironment{longlisting}{\captionsetup{type=listing}}{}

\theoremstyle{definition}

\theoremstyle{definition}
\newtheorem{definition}{Definition}[section]
\newtheorem{example}{Example}[section]
\newtheorem*{remark}{Remark}

\begin{document}

\title{Normalization, Square Roots, and the Exponential and Logarithmic Maps in Geometric Algebras of Less than 6D}

\author{Steven De Keninck}
\orcid{0000-0002-6870-1714}
\authornotemark[1]
\affiliation{%
  \department{Informatics Institute}
  \institution{University of Amsterdam}
  \city{Amsterdam}
  \country{The Netherlands}
}
\email{steven@enki.ws}
\author{Martin Roelfs}
\authornote{Both authors contributed equally to the paper}
\orcid{0000-0002-8646-7693}
\affiliation{%
  \department{Department of Physics}
  \institution{KU Leuven}
  \city{Kortrijk}
  \country{Belgium}
}
\email{martin.roelfs@kuleuven.be}

\begin{abstract}
Geometric algebras of dimension $n < 6$ are becoming increasingly popular for the modeling of 3D and 3+1D geometry. With this increased popularity comes the need for efficient algorithms for common operations such as normalization, square roots, and exponential and logarithmic maps. The current work presents a signature agnostic analysis of these common operations in all geometric algebras of dimension $n < 6$, and gives efficient numerical implementations in the most popular algebras \GR{4}, \GR{3,1}, \GR{3,0,1} and \GR{4,1}, in the hopes of lowering the threshold for adoption of geometric algebra solutions by code maintainers.
\end{abstract}

\maketitle
\renewcommand{\shortauthors}{Roelfs and De Keninck}

    \let\svthefootnote\thefootnote
    \let\thefootnote\relax\footnotetext{
    Authors’ addresses: Steven De Keninck, Informatics Institute,
    University of Amsterdam, Amsterdam, The Netherlands, steven@enki.ws; Martin Roelfs, Department of Physics, KU Leuven, Kortrijk,
    Belgium, martin.roelfs@kuleuven.be.
    }
    \let\thefootnote\svthefootnote


\section{Introduction}

Low dimensional real geometric algebras, ranging from the 3 dimensional vectorspace algebra $\GR{3}$; to the 4 dimensional algebras $\GR{4}$, spacetime algebra $\GR{3,1}$, and Euclidean PGA $\GR{3,0,1}$; up to the 5 dimensional conformal $\mathbb R_{4,1}$, are becoming increasingly popular to model the various geometries of 3D space and 3+1D spacetime. Compared to the more traditional vector/matrix approaches they often offer substantial simplifications and generalisations, freeing the user from the burden of dealing with coordinate-based notations for representations, co/contra-variance, and much more.

With this increased popularity comes the need for efficient and stable numerical methods for the manipulation of the multivectors that replace matrices and vectors. In this paper we introduce implementations for the most commonly used of these methods. 

First we turn our attention to the renormalization of rotors, the multivector equivalent of the Gram–Schmidt procedure or SVD used to reorthogonalize matrices when e.g. numerical integration makes a group element drift away from the motion manifold. 
We show how this new normalization procedure trivializes the calculation of square roots, an operation that, because of the frequent use of conjugation in geometric algebra, is much more common than it is when working with matrices.  
Next, we turn our attention to 
the exponential and logarithmic maps of rotors, and show how the general procedure in any number of dimensions \cite{roelfs2021graded}, can be simplified in each of these specific low dimensional cases. 
Finally we present optimized implementations that we hope will lower the threshold for code maintainers to transition to a geometric algebra based code base.

A signature and dimension agnostic approach to the exponential and logarithmic maps for rotors in all geometric algebras \GR{p,q,r} was recently published \cite{roelfs2021graded}. 
However, the topic of renormalization has so far experienced disparate treatment depending on the signature of the algebra \cite{STA,GunnThesis,DorstDecomposition,GeometricCalculus}, and is limited to geometric algebras of dimension $n < 6$. 
The limit at $n = 6$ occurs because at this point quadvectors stop squaring to scalars, a property these methods take advantage of. (An example of this breakdown is the quadvector $\e{1234} + \e{3456}$ in \GR{6}, which squares to $2 + 2\e{1256}$.)
However, by applying the approach set out in \cite{roelfs2021graded} to the existing literature on normalization \cite{STA,GunnThesis,DorstDecomposition,GeometricCalculus}, a signature agnostic normalization procedure in geometric algebras of $n < 6$ was found.
While the $n \geq 6$ limit remains intractable, the signature agnostic approach laid out in the current work provides additional insights that might guide future work towards $n=6$ and beyond.
Fortunately however, many popular algebras are of dimension $n < 6$.

This paper is organized as follows. \Cref{sec:three_dimensional} briefly describes how all of the previously mentioned algebras can be used to represent various 3D geometries in order to give an intuitive understanding. \Cref{sec:kreflections} describes how all isometries are formed by the composition of reflections. \Cref{sec:squareroot} conceptually outlines the square roots of continuous isometries. \Cref{sec:renormalization} describes the renormalization algorithm that enables efficient renormalization of a multivector that has drifted away from the rotor manifold and shows how this can be used for the calculation of square roots. \Cref{sec:exponential} presents the method by which the exponential of any bivector can be calculated. \Cref{sec:logarithm} present the logarithmic map. Finally, \cref{sec:numerical} gives efficient numerical algorithms for these operations in the previously mentioned algebras.

\section{Three Dimensional Geometries}\label{sec:three_dimensional}
In order to describe geometry in three dimensional space \ER{3}, various geometric algebras exist. 
The simplest is \GR{3}, the geometric algebra of three basis vectors $\e{i}$ satisfying $\e{i}^2 = 1$. Because a plane through the origin is described by a linear equation, vectors in \GR{3} can directly be used to represent planes through the origin:
    \[ u_1 x_1 + u_2 x_2 + u_3 x_3 = 0 \quad \rightarrow \quad u = u_1 \e{1} + u_2 \e{2} + u_3 \e{3}. \]
Two planes $u$ and $v$ intersect in the line through the origin $u\wedge v$, and three planes $u$, $v$ and $w$ intersect in the origin $u \wedge v \wedge w$ itself.
While it is possible to describe all of three dimensional geometry with this origin attached model, it requires semantic overloading of the limited set of elements and operations (e.g. direction vectors vs. location vectors), leading to quite involved non-algebraic administration of data types.
However, consider the equation for a general plane in 3D:
    \[ u_1 x_1 + u_2 x_2 + u_3 x_3 + u_0 = 0. \]
This can be represented as a vector in a space of one dimension higher by adding one additional basis vector $\e{0}$ to represent the plane at infinity, a.k.a. the horizon, such that the plane can be represented by the vector
    \[ u = u_0 \e{0} + \sum_{j=1}^3 u_j \e{j}. \]
There are three choices for the signature of this extra basis vector $\e{0}$: $\e{0}^2 \in \{ 1, 0, -1 \}$. These lead to elliptic, Euclidean, and hyperbolic projective geometry respectively, and are realized using the geometric algebras \GR{4}, \GR{3,0,1} and \GR{3,1} \cite{GunnThesis}.
In order to describe Euclidean 3D space and the isometries therein, the Euclidean Projective Geometric Algebra \GR{3,0,1} with its additional null vector is the correct choice, as (compositions of) reflections in this space form the Euclidean group \Eu{3}.
The fact that the isometries of \GR{3,0,1} correspond so directly to the Euclidean group \Eu{3} is both due to the fact that the horizon $\e{0}$ is a null vector, and due to the identification of vectors with planes rather than points.
The elements of the Euclidean group \Eu{3} are compositions of reflections in planes, with two reflections composing into rotations \& translations, three reflections forming improper rotations \& translations, and four reflections combining into screw motions.

Conformal transformations are obtained when we replace reflections in planes by inversions in spheres. In order to do this, two extra basis vectors $\e{+}$ and $\e{-}$ satisfying $\e{+}^2 = -\e{-}^2 = 1$ in the geometric algebra \GR{4,1} are used to form two null vectors in a Witt basis:
    \[ n_o := \tfrac 1 2 (\e{-} + \e{+}), \quad n_\infty := \e{-} - \e{+}.
    \]
Here $n_o$ is a sphere of zero radius centered on the origin, and $n_\infty$ is the sphere at infinity.
Now a sphere of radius $\rho$ centered on $x = x_1 \e1 + x_2 \e2 + x_3 \e3$ is represented by the vector
    \[ u = n_o + x + \tfrac 1 2 (x^2 - \rho^2) n_\infty. \]
All isometries in these various 3D geometries can be represented using products of vectors, which leads us to the notion of $k$-reflections.

\section{$k$-Reflections}\label{sec:kreflections}
\begin{figure*}[htb]
    \centering
    \includegraphics[width=\textwidth]{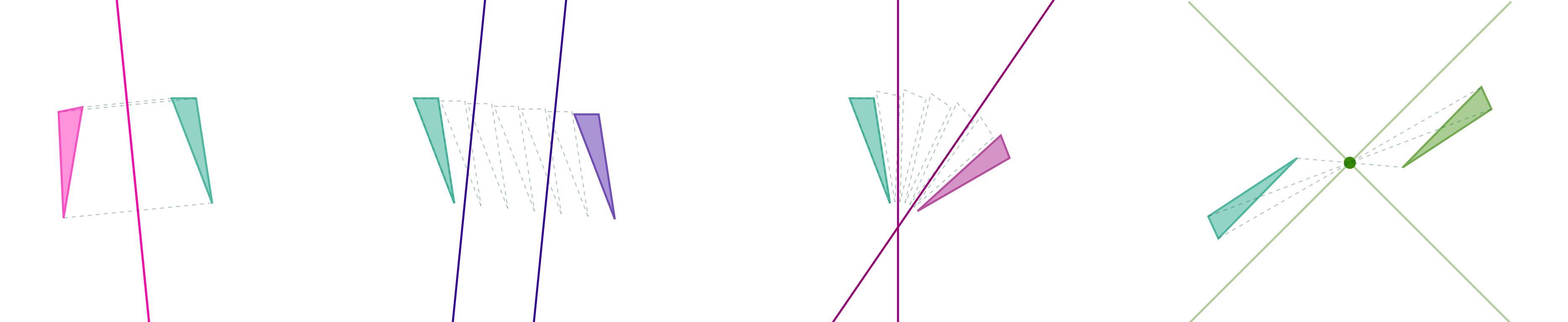}
    \caption{Reflections in hyperplanes make up all the transformations \emph{and} elements of Euclidean Geometry. From left to right, a line-reflection (which uniquely represents that line), two parallel line-reflections compose into a translation, two intersecting line-reflections constitute a rotation, and two orthogonal line-reflections form a point-reflection (which uniquely represents that point).}
    \label{fig:reflections}
\end{figure*}
All isometries and elements of geometry such as points, lines, planes, in an $n$ dimensional space, are compositions of at most $k$ reflections \cite{roelfs2021graded}.
As an illustration in 2D geometry, \cref{fig:reflections} shows how two reflections can compose into either a translation when the two reflections are parallel and intersect at infinity, a rotation when the two reflections intersect in the space itself, and a point reflection in the special case when the two reflections are orthogonal.
All isometries can therefore be described as $k$-reflections, i.e. the composition of $k$ non-collinear reflections.

A hyperplane $v$ is reflected in the hyperplane $u$ by the conjugation rule
    \[ v \mapsto u[v] := -u v u^{-1}. \]
The minus sign in the conjugation rule ensures that a hyperplane $v$ reflected in itself switches the direction it faces: $v[v] = -v$. Because performing the same reflection twice should be the same as doing nothing, we find from $u[u[v]] = v$ that $u^2 = \pm 1$.\footnote{In fact, the only requirement is for $u^2$ to be a non-zero scalar. The normalization is merely a convenient choice.}
\begin{definition}[Reflection]
A normalized hyperplane $u$ satisfying $u^2 = \pm 1$ is called a reflection.
\end{definition}
It might seem like this definition confuses the element with the act of reflecting. However, this is deliberate, and comes from the philosophical standpoint to directly identify elements of geometry with transformations, a mindset we hope to have conveyed by the end of this section.

Repeated reflections in $k \leq n$ non-collinear reflections produce a $k$-reflection $U = u_1 u_2 \cdots u_k$. The conjugation rule for $U$ being applied to an $l$-reflection $V = v_1 v_2 \ldots v_l$ is found by demanding that it transforms covariantly:
    \begin{equation}
        U[ v_1 v_2 \ldots v_l] = U[v_1] U[v_2] \cdots U[v_l].
    \end{equation}
I.e. transforming $V$ is identical to transforming each of its constituent reflections $v_i$ in turn. This yields the conjugation rule \cite{roelfs2021graded}
    \begin{equation}\label{eq:conjugation}
        V \mapsto U[V] = (-1)^{\grade(U) \grade(V)} U V U^{-1},
    \end{equation}
where the grade of a $k$-reflection is $k$.

\begin{figure*}[htb]
    \centering
    \includegraphics[width=\textwidth]{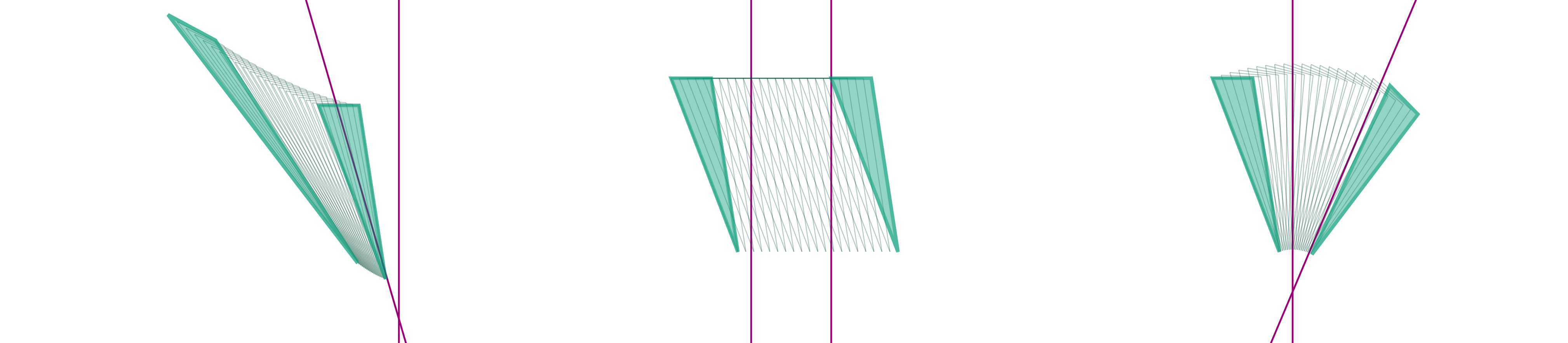}
    \caption{The three basic types of bireflections. Boosts, translations and rotations. (left) Any boost is isomorphic to \Spin{1,1}, with the intersection of the two reflections as its origin. (middle) Any translation is isomorphic to \Spin{1,0,1}, with the intersection point of the two reflections on the horizon as its origin.
    (right) Any rotation is isomorphic to \Spin{2}, with the intersection point of the two reflections as its origin.}
    \label{fig:bireflections}
\end{figure*}

\begin{definition}[Bireflection]
    A bireflection $R$ is the composition of two reflections $u$ and $v$: 
        \[ R = vu, \]
    and under conjugation performs a reflection in $u$ followed by a reflection in $v$.
    In case $vu = v \wedge u$, i.e. the two reflections $u$ and $v$ are orthogonal, the bireflection can be used to represent an element of geometry. 
    For example, in 2D Euclidean space reflections in two orthogonal lines cause a point reflection, and thus the bireflection $vu$ can be used to represent that point (\cref{fig:reflections}).
\end{definition}
\noindent
A bireflection $R=vu$ satisfying $u^2 = v^2$ also satisfies the \emph{rotor condition}
    \begin{equation}
        R \widetilde{R} = 1,
    \end{equation}
and because it is the composition of two reflections it is called a \emph{simple rotor} \cite{GeometricCalculus}.
The two hyperplanes $u$ and $v$ intersect in a hyperline $\bl \propto v \wedge u$.
In an $n$ dimensional space, a hyperplane defines an $n-1$ dimensional subspace, and thus the hyperline $\bl$ specifies an $n-2$ dimensional subspace, which is invariant under the action of the bireflection $vu$.
Therefore, any simple rotor is isomorphic to either $\Spin{2} \cong \Spin{0,2}$ when it generates a rotation around a hyperline $\bl$, isomorphic to $\Spin{1,1}$ if it generates a boost along a hyperline $\bl$, or isomorphic to $\Spin{1,0,1} \cong \Spin{0,1,1}$ if it generates a translation in the direction orthogonal to the infinite hyperline $\bl$.
These scenarios are shown in \cref{fig:bireflections} for 2D.
Because $\bl$ squares to a scalar $\lambda := \bl^2 \in \mathbb{C}$, it is called a \emph{simple} bivector.\footnote{For rotations $\lambda < 0$, for boosts $\lambda > 0$, and for translations $\lambda = 0$. Therefore, these common continuous isometries all satisfy $\lambda \in \mathbb{R}$. However, more exotic scenarios where $\lambda \in \mathbb{C}$ do occur, particularly in the invariant factorization of certain rotors \cite{roelfs2021graded}. Their geometric interpretation is an open question. \label{ftn:exotic}}
By repeated multiplication the bireflection $R$ defines a one parameter subgroup of bireflections $R(\theta) = R^\theta = e^{\theta \log(R)}$, where $\log(R) = \bl$, and thus any bireflection can be written as
    \begin{align}\label{eq:euler}
        R &= e^{\bl} = \sum_{j=0}^\infty \tfrac{1}{j!} \bl^{j} \\
        &= \sum_{j=0}^\infty \tfrac{1}{(2j)!} \lambda^{j} + \bl \sum_{j=0}^\infty \tfrac{1}{(2j+1)!} \lambda^{j}, \label{eq:euler_line2}
    \end{align}
where the two sums in \cref{eq:euler_line2} are scalar quantities.
All bireflections therefore follow Euler's formula, if we define
\begin{alignat}{7}
        \co\pqty{\bl} &:= u \cdot v &&= \expval{R} &&= \tfrac{1}{2} \pqty{e^{\bl} + e^{-\bl}}
        &&= \cosh(\sqrt{ \bl^2 }) 
        \label{eq:generalized_cos} \\
        \si\pqty{\bl} &:= u \wedge v &&= \expval{R}_2 &&= \tfrac{1}{2} \pqty{e^{\bl} - e^{-\bl}} 
        &&= \bl \sinhc\pqty{\sqrt{ \bl^2 }},
        \label{eq:generalized_sin}
    \end{alignat}
where the $\sinhc$ function is the hyperbolic analog of the $\sinc$ function, defined for $z \in \mathbb{C}$ by
    \begin{equation}
        \sinhc(z) := \sum_{j=0}^\infty \tfrac{1}{(2j+1)!} z^{2j} = \begin{cases}
            \frac{\sinh(z)}{z} & z \neq 0 \\
            1 & z = 0
        \end{cases} .
    \end{equation}
When $\lambda = - 1$, \cref{eq:euler} reduces to the familiar Euler's formula for complex numbers.
But because the $\cosh$ and $\sinh$ functions are well defined on the entire complex plane, \cref{eq:generalized_cos,eq:generalized_sin} cover rotations, translation and boosts alike.\footref{ftn:exotic}
Further defining the generalized tangent function $\ta(\bl) = \si(\bl)/ \co(\bl)$, the following are equivalent methods for calculating the exponential of a simple bivector $\bl$:
    \begin{equation}
        R = \co(\bl) + \si(\bl) = \co(\bl) \bqty{1 + \ta(\bl)}.
    \end{equation}
\begin{definition}[Trireflection]
    A trireflection $P$ is the composition of three reflections $u$, $v$ and $w$: 
        \[ P = wvu. \]
    Any trireflection can be decomposed into a commuting reflection $r$ and bireflection $R$ such that $P = R r = r R$. If the vector part $\expval{P}_1 \neq 0$, the solution is uniquely given by
        \[ r = \overline{\expval{P}_1}, \qquad R = r^{-1} P = P r^{-1}, \]
    where $\overline{X}$ denotes normalization such that $r^2 = \pm 1$.
    However, if $\expval{P}_1 = 0$ the solution is no longer unique. This occurs because all three reflections are orthogonal, and thus $wvu = w \wedge v \wedge u$. However, in this case it suffices to calculate the projection $R = \overline{x \cdot P}$ for any vector $x$ satisfying $x \cdot P \neq 0$, after which $r = P R^{-1} = R^{-1}P$, so a decomposition into commuting factors always exists.
    
    A trireflection of orthogonal reflections doubles as an element of geometry. 
    For example, in 3D Euclidean space three reflections in orthogonal planes cause a point reflection, and thus the trireflection $wvu$ can be used to represent that point.
\end{definition}
Application of a trireflection $P = Rr$ in 3D space can be explained as follows: first, the entire space is reflected in $r$. This leaves only the plane $r$ itself invariant. This plane spans a 2DPGA subspace, in which we can still apply a bireflection $R$, which can be either a rotation or a translation. Together, the three reflections then form improper rotations or glide reflections respectively. In the special case when all three reflections are orthogonal, the resulting isometry is a point reflection in the intersection point $w \wedge v \wedge u$ of the three reflections.

\begin{figure*}[htb]
    \centering
    \includegraphics[width=\textwidth]{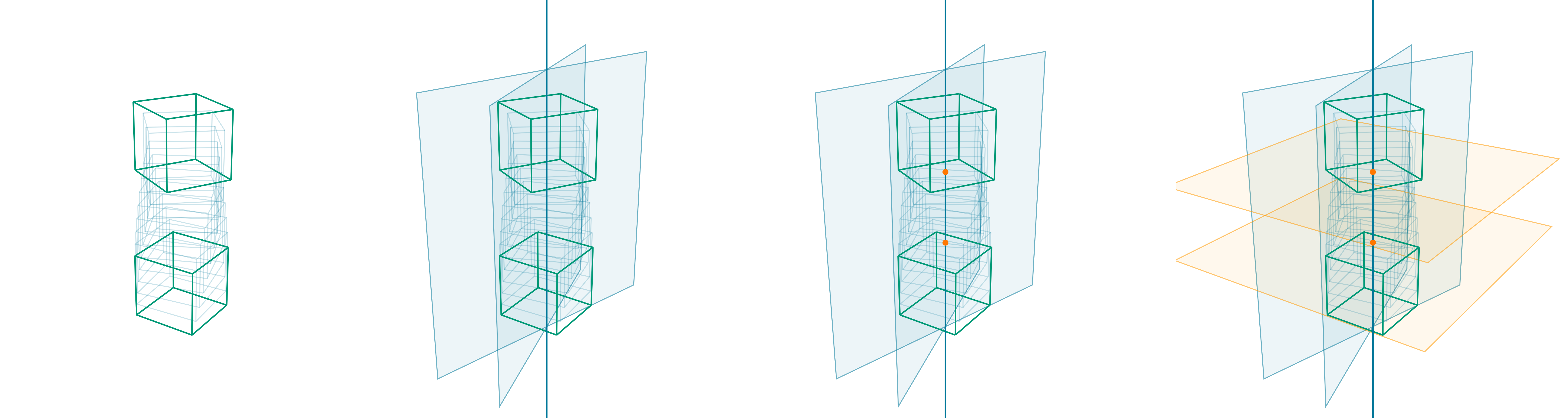}
    \caption{(a) The cube is undergoing a screw motion. (b) It is rotated around the rotational axis of a bireflection. This axis is invariant under the rotation. (c) The axis is a 1D subspace in which there is only one continuous transformation left: translation along the line. The translation is again the result of a bireflection: two point reflections. (d) The point-reflections generating the translation lie on planes which are also invariant under the rotation. Hence, the total screw motion is the composition of four reflections in planes. }
    \label{fig:chasles}
\end{figure*}

\begin{definition}[Quadreflection]\label{def:quadreflection}
    A quadreflection $R$ is the composition of four reflections $a$, $b$, $c$ and $d$: 
        \[ R = dcba. \]
    Any quadreflection can be decomposed into two commuting bireflections $R_1 = v_1 u_1$ and $R_2 = v_2 u_2$, such that
        \[ R = R_2 R_1 = R_1 R_2. \]
    The decomposition of $R$ into $R_1$ and $R_2$ is essentially given in \cref{sec:logarithm}, for more details see \cite{roelfs2021graded}.
\end{definition}
In 3D Euclidean space, at most four reflections can be combined to form a quadreflection, commonly revered to as a screw-motion, which is shown in \cref{fig:chasles}.
The first bireflection leaves a single line invariant: the axis of rotation. This line is a 1DPGA subspace, and thus the only bireflection left that can still be performed within this subspace is the combination of two point reflections, which generates a translation. When viewed in the full 3D space these points extend out into planes which are left invariant under the rotation. Thus, the screw motion is the composition of four reflections. This is the famous Mozzi-Chasles' theorem, which states that the most general continuous isometry in 3D can be decomposed into a commuting rotation and translation.

\subsection*{Elements of Geometry}
Elements of geometry such as planes, lines, spheres etc., are compositions of $k$ orthogonal reflections and therefore also $k$-reflections. In 3D geometry a single reflection represents a plane (or sphere in 3DCGA). Two reflections intersect in a line (or circle in 3DCGA), and thus a line can be represented by a bireflection $vu$ of two orthogonal reflections such that $vu = v \wedge u$. 
Three reflections intersect in a point (or a point-pair in 3DCGA), and thus a point can be represented by a trireflection $wvu$ of three orthogonal reflections such that $wvu = w \wedge v \wedge u$.
The identification is supported by the fact that reflecting in each of the orthogonal planes $w$, $v$ and $u$ in turn, is identical to a reflection in the point $wvu$.
Thus, geometric algebra puts isometries and elements of geometry on equal footing as $k$-reflections.

\section{Square root of a $2k$-reflection}\label{sec:squareroot}
Consider two normalized elements of geometry $A$ and $B$ of identical type, i.e. $A$ and $B$ are both $k$-reflections of orthogonal reflections and thus normalized blades. The rotor from $A$ to $B$ is given by the ratio $R = B / A$, since $(B/A)A = B$. However, because $k$-reflections are applied using the two sided conjugation rule \cref{eq:conjugation}, we instead need to find $\sqrt{B/A}$, such that it can be applied using conjugation as $\pqty{B/A}^{1/2} A \pqty{B/A}^{-1/2}$. 
This might seem unnecessarily complicated, but is absolutely necessary if the same rotor is to be applied to other elements of the algebra to rotate the entire scene from $A$ to $B$.
This makes the square root of a $2k$-reflection a very commonly used operation in geometric algebras.
\begin{figure*}[htb]
    \centering
    \includegraphics[width=\textwidth]{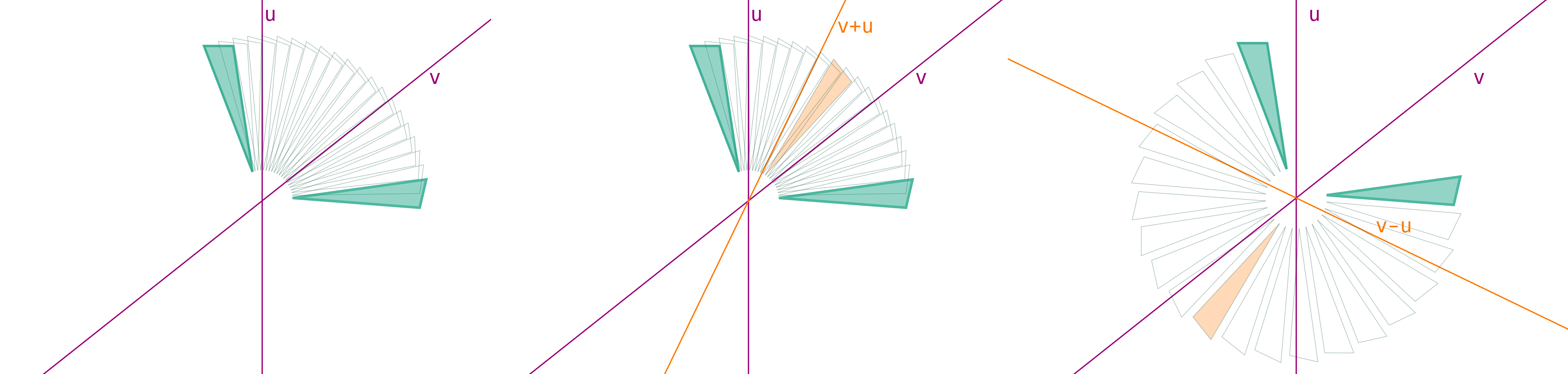}
    \caption{The bireflection $vu$ generates a rotation over twice the angle at which $v$ and $u$ intersect by first reflecting in $u$, and then in $v$. In order to generate only half the rotation, we instead need to reflect in $u$, and then $\overline{v \pm u}$. These generate two different final states, exactly $\pi$ radians apart.}
    \label{fig:sqrt_R}
\end{figure*}

For a geometric understanding of the square root, consider the simple case when $R = vu$ is a bireflection generating a rotation in the plane \GR{2}, as shown in \cref{fig:sqrt_R}. Geometrically, we see that to perform half the rotation of $vu$, we should instead reflect over $u$ and a bisector $\overline{v \pm u}$, where $\overline{v \pm u}$ denotes normalization such that $\overline{v \pm u}^2 = \pm 1$ depending on the signature of the space. Thus, we know that $\sqrt{vu} = (\overline{v \pm u}) u$, and we find that
\begin{align}
    \sqrt{vu} \propto (v \pm u) u = \pm 1 + vu.
\end{align}
The bireflection $vu$ has two distinct square roots corresponding to the two bisectors $\overline{v \pm u}$, since we can rotate towards the final state in either direction.
The natural notion of the principal square root, which carries over to translations and boosts, is to take the inner bisector $\overline{u + v}$, and thus we define the principal square root of any $2k$-reflection as
\begin{equation}
    \sqrt{R} := \overline{1 + R}.
\end{equation}
Only considering this principal square root leads to a rotor representation of \SO{2}, the half-cover of \Spin{2}, as is depicted in \cref{fig:sqrt_R}.
\begin{remark}
This definition of the principal square root is the only definition that is valid for rotations, translations, and boosts alike.
First, for translations the vectors $u$ and $v$ are parallel and thus the bisector $v - u = \e{0}$ is always the horizon $\e{0}$, and one cannot reflect over the horizon, something mathematically encapsulated in the property $\e{0}^2 = 0$.
Thus, for translations the only valid bisector is the average $\overline{u + v}$.
Second, for boosts the two reflections $u$ and $v$ both have to be timelike, i.e. $u^2 = v^2$. The average bisector $\overline{u + v}$ is also timelike, because $\pqty{\overline{u + v}}^2 = u^2$.
Contrarily, the bisector $\overline{v - u}$ is spacelike because $\pqty{\overline{v - u}}^2 = -u^2$, and thus it does not lie in the same connected component of \Spin{1,1} as $u$. 
So although $(\overline{v - u}) v$ is still a square root of $vu$, it is not connected to $vu$ via a continuous transformation.
\end{remark}

For the two possible square roots of a simple rotor, the normalization constant is either the scalar quantity $2 \co(\bl)$, or the bivector quantity $2 \si(\bl)$, since
    \[ X_\pm = \pm 1 + R = \pqty{\pm R^{-1/2} + R^{1/2}} R^{1/2}
    = \begin{cases}
        2 \co(\bl) \sqrt{R} & \text{if }+ \\
        2 \si(\bl) \sqrt{R} & \text{if }- \\
    \end{cases}. 
    \]
The scalar quantity $2 \co(\bl)$ can uniquely be determined from $X_+ \widetilde{X}_+ = 4 \co^2(\bl)$ up to sign, whereas there is no unique way to obtain $2 \si(\bl)$ from $X_+ \widetilde{X}_+ = 4 \si^2(\bl)$ since there is an infinity of bivectors which square to $\si^2(\bl)$. This provides further motivation for the definition of $\sqrt{R} = \overline{1+R}$ as the principal square root.

Now we consider the principal square root of a quadreflection $R = dcba$ in $n < 6$ (\cref{def:quadreflection}), which additionally satisfies the rotor condition $R \widetilde{R} = 1$. Such a quadreflection is generated by a non-simple bivector $\B$ and thus $\B \wedge \B \neq 0$. 
Nonetheless, $\B$ is the sum of two orthogonal commuting simple bivectors $\bl_1$ and $\bl_2$ which do square to scalars $\lambda_i = \bl_i^2 \in \mathbb{C}$ \cite{roelfs2021graded}, and thus $\B$ squares to
    \[ \B^2 = \lambda_1 + \lambda_2 + 2 \bl_1 \bl_2. \]
The quadreflection can thus be written as $R = e^\B = e^{\bl_1} e^{\bl_2}$, a property which will be essential to the computation of the logarithm of $R$ in \cref{sec:logarithm}.
We can still define the generalized cosine and sine series for a quadreflection as
    \begin{align}
        \co(\B) &:= \tfrac{1}{2} \pqty{R + \widetilde{R}} = \co(\bl_1)\co(\bl_2) + \si(\bl_1)\si(\bl_2) \\
        \si(\B) &:= \tfrac{1}{2} \pqty{R - \widetilde{R}} = \si(\bl_1)\co(\bl_2) + \co(\bl_1)\si(\bl_2).
    \end{align}
Thus, $\si(\B)$ still satisfies $\widetilde{\si(\B)} = -\si(\B)$ and is still a bivector, while $\co(\B)$ still satisfies $\widetilde{\co(\B)} = \co(\B)$ \emph{but} has become scalar plus quadvector.
We now apply our definition for the principal square root as $1 + R$ to the quadreflection $R$ and find
    \[ X = 1 + R = (R^{-1/2} + R^{1/2}) R^{1/2} = 2\co(\B / 2) \sqrt{R}  \; \implies \; \sqrt{R} = \frac{X}{2 \co(\B/2)}. \]
Thus, $X$ is still identical to $2\co(\B / 2) \sqrt{R}$, but the proportionality constant $2 \co(\B / 2)$ has become scalar plus quadvector, and $\sqrt{R}$ can still be calculated from $1 + X$ upon renormalization if the renormalization procedure correctly handles non-scalar norms.
This necessarily brings us to the notion of renormalization for non-scalar norms.

\section{Renormalization, Study Numbers \& Polar  Decomposition}\label{sec:renormalization}
In geometric algebras of dimension $n \leq 3$, the quantity $X \widetilde{X}$ is always a scalar. It is therefore standard practice to define the norm of an element $X$ as the scalar quantity
    \begin{equation}
        \norm{X} := \sqrt{\vert X \widetilde{X} \vert}.
    \end{equation}
With this norm, any element $X$ can be normalized as
    \begin{equation}\label{eq:classicnormalization}
        \overline{X} = \frac{X}{\norm{X}} = \frac{X}{\sqrt{ \vert X \widetilde{X} \vert }}.
    \end{equation}
However, when $n > 3$, the quantity $X \widetilde{X}$ is no longer guaranteed to be a scalar.
A common solution is to only consider the scalar part of $X \widetilde{X}$ and instead to define the norm as $\norm{X} := {\vert \langle X \widetilde{X} \rangle \vert}^{1/2}$, but this discards an important part of the story.
All we know for certain is that the norm is a self-reverse quantity, since 
\[ \widetilde{X \widetilde{X}} = X \widetilde{X}. \]
Consequentially, $X \widetilde{X}$ contains only terms of grade $k$, where $0 \cong k \mod{4}$ or $1 \cong k \mod{4}$.
Defining $S^2 := X \widetilde{X}$ with $S$ a self-reverse number, $X$ permits a polar decomposition into $X = S R$ \cite{DorstDecomposition}, where $R \in \Pin{p,q,r}$ is a $k$-reflection. 
Thus, $S = \sqrt{X \widetilde{X}}$, and $R = S^{-1} X$. 
In general no closed form computation of $S^{-1} = (X \widetilde{X})^{-1/2}$ is known, but in the particular case where
    \[ X \widetilde{X} = \langle X \widetilde{X} \rangle + \langle X \widetilde{X} \rangle_4, \]
with $\langle X \widetilde{X} \rangle_4^2 \in \mathbb{R}$, the computation of $S^{-1} = (X \widetilde{X})^{-1/2}$ is straightforward. 
The condition $\langle X \widetilde{X} \rangle_4^2 \in \mathbb{R}$ is always met when $X$ is an element of an even subalgebra $\GR[+]{pqr}$ with $p+q+r < 6$, but seizes to be universally true when $n \geq 6$.
Numbers satisfying this condition are called Study numbers \cite{GunnThesis}.

\subsection{Study Numbers}

A multivector is called a Study number $S$ when it can be split into a sum
    \begin{equation}
        S = a + b I,
    \end{equation}
of a scalar part $a \in \mathbb R$ and a non-scalar part $bI \in \GR{p,q,r}$ whose square is a scalar: $(bI)^2 \in \mathbb R$.
When $(bI)^2 < 0$, the Study number is isomorphic to a complex number, when $(bI)^2 > 0$ it is isomorphic to a split-complex number, and when $(bI)^2 = 0$ it is isomorphic to a dual number. 
In \GR[+]{p,q,r} for $n < 6$, $I$ is the pseudoscalar of the space, but in general this notation is meant as a mnemonic for these isomorphisms.
For example, for $n=4$ and $n=5$, $I$ is the pseudoscalar of the algebra \GR[+]{p,q,r}, so that $b$ is a scalar for $n=4$, and a vector for $n=5$.
As a number system we denote the Study numbers as $\mathbb{S}$.

We have already seen one example of Study numbers: bireflections $vu$ are Study numbers, since $\langle vu \rangle~\in~\mathbb{R}$ and $\langle vu \rangle_2^2 \in \mathbb{R}$. 
However, the most important for the purpose of this work are Study numbers in $n < 6$ satisfying $S = \langle S \rangle + \langle S \rangle_4$.
Because such Study numbers are self-reverse, their conjugate cannot be found in terms of the standard involutions of geometric algebra. So instead, we define the conjugate of a Study number as
    \begin{equation}
        \breve{S} := a - bI
    \end{equation}
Analogously to e.g. complex numbers, the norm of a Study number is the scalar value
    \begin{equation}\label{eq:norm_S}
        \norm{S}_\mathbb{S} := \sqrt{S \, \breve{S}} = \sqrt{a^2 - (bI)^2}.
    \end{equation}
Using this norm, the inverse of a Study number $S$ is given by
    \begin{align}
        S^{-1} = \frac{\breve{S}}{\norm{S}_\mathbb{S}^2} = \frac{a - bI}{a^2 - (bI)^2}.
    \end{align}

\subsection{Square Root of a Study Number}\label{sec:squareroot_S}
Because $(bI)^2 \in \mathbb{R}$, the square root, or indeed any power of a Study number, is again a Study number. Therefore, we know that the square root of $S = a + bI$ is itself a Study number $\sqrt{S} = \sqrt{a + b I} = c + dI$. Solving $a + bI = (c + dI)^2$ allows us to find a closed form solution for the square root of a Study number:
    \begin{align}
        a + bI = (c^2 + d^2 I^2) + (2 cd I).
    \end{align}
Solving for $c$ and $dI$ yields
    \begin{equation}
        c_\pm = \sqrt{\frac{a \pm \sqrt{a^2 - b^2 I^2}}{2}}, \qquad d I = \frac{b}{2c_\pm}I.
    \end{equation}
with which $\sqrt{S}$ can be expressed purely in terms of $S$ as
    \begin{equation}\label{eq:sqrt_study}
        \sqrt{S} = c_\pm + \frac{1}{2c_\pm} bI, \qquad c_\pm = \sqrt{\tfrac{1}{2} \pqty{\expval{S} \pm \norm{S}_\mathbb{S}}}.
    \end{equation}
For future convenience, we can now additionally calculate $S^{-1/2}$ as
    \begin{equation}
        S^{-1/2} = \frac{4 c_\pm^3}{4 c_\pm^4 - (bI)^2} - \frac{2 c_\pm}{4 c_\pm^4 - (bI)^2} bI.
    \end{equation}
Clearly, $S^{-1/2}$ is well defined when the denominator $4 c_\pm^4 - (bI)^2 \neq 0$, and hence $S^{-1/2}$ does not exist when 
\[ \norm{S}_\mathbb{S} (\norm{S}_\mathbb{S} \pm \expval{S}) = 0. \]
Ergo, $S^{-1/2}$ does not exist either when the Study number $S$ is singular and thus $\norm{S}_\mathbb{S} = 0$, or when $\expval{S} = \mp \norm{S}_\mathbb{S}$ depending on the choice for $c_\pm$. Thus the choice of $c_\pm$ amounts to the choice of where to place the branch cut. 
When $I^2 = -1$, the natural choice is to use $c_+$, which places the branch cut on the negative real axis and coincides with the traditional choice for the branch cut of $\sqrt{z}$ for $z \in \mathbb{C}$ made in complex analysis. However, when $\expval{S} = - \norm{S}_\mathbb{S}$, the other square root determined by $c_-$ is still valid. 
Hence, for the square root of a negative real number we use $c_-$. 
An example is given in \cref{ex:bivectors}.
When $I^2 = +1$, any Study number $S = a \pm aI$ has $\norm{S}_\mathbb{S} = 0$, which we recognize as the light-cone (or null-cone), and thus a space with $I^2 = 1$ has two additional branch cuts.

\subsection{Renormalization}
We now return to the renormalization of a general $X$  in the even subalgebra $\GR[+]{p,q,r}$, 
using the polar decomposition into $X = SR$ as described in \cref{sec:renormalization}. 
Observing that $S^2 = X \widetilde{X}$ is a self-reverse Study number with $a = \langle X \widetilde{X} \rangle$ and $bI = \langle X \widetilde{X} \rangle_4$, we find
    \begin{align}\label{eq:normalization}
        R = \pqty{X \widetilde{X}}^{-1/2} X &= \frac{4 c_\pm^3}{4 c_\pm^4 - \langle X \widetilde{X} \rangle_4^2} X - \frac{2 c_\pm \langle X \widetilde{X} \rangle_4}{4 c_\pm^4 - \langle X \widetilde{X} \rangle_4^2} X, \quad 
        \text{where }c_\pm = \sqrt{\tfrac{1}{2} ( \langle X \widetilde{X} \rangle + \Vert X \widetilde{X}\Vert_\mathbb{S})}.
    \end{align}
As shown in \cref{sec:squareroot_S}, at least one square root $(X \widetilde{X})^{-1/2}$ exists unless $\Vert X \widetilde{X}\Vert_\mathbb{S} = 0$, i.e. when $X \widetilde{X}$ is singular.
The rotor obtained from \cref{eq:normalization} satisfies the rotor condition $R \widetilde{R} = 1$, and is thus an orthogonal transformation \cite{GA4CS}, meaning that the transformation $y \to R y \widetilde{R}$ preserves the inner product $u \cdot v$ between vectors $u, v \in \GR[(1)]{p,q,r}$:
    \begin{align}\label{eq:inner_product}
        (R u \widetilde{R}) \cdot (R v \widetilde{R}) &= \tfrac{1}{2} \pqty{ R u \widetilde{R} R v \widetilde{R} + R v \widetilde{R} R u \widetilde{R}} \\
        &= \tfrac{1}{2} R \pqty{u v  + v u } \widetilde{R} = R(u \cdot v)\widetilde{R} = u \cdot v.
    \end{align}
This normalization procedure replaces traditional reorthogonalization procedures based on Singular Value Decomposition (SVD) or the (modified) Gram-Schmidt orthogonalization by a numerically efficient method based purely within GA.
    
Note that the conjugation formula \cref{eq:conjugation} uses inversion rather than reversion, which follows from $R^{-1} = \widetilde{R}$. 
However, \emph{any} invertible element $X \in \GR[+]{p,q,r}$ preserves the inner product of a vector $y$ under the transformation $y \to X y X^{-1}$, following an argument similar to that of \cref{eq:inner_product}. 
But by the polar decomposition $X = SR$ all of these transformations are proportional to a rotor $R$, up to a Study number $S$.
The defining quantity of the transformation is thus $R$, which allows the inverse in the conjugation formula \cref{eq:conjugation} to be replaced by the more economical reverse. 
Therefore it could be argued that \emph{orthonormal} transformation would be a more fitting name for $R y \widetilde{R}$ than the more commonly used \emph{orthogonal} transformation.

\begin{remark}
It is noteworthy that \emph{any} element $X \in \GR[+]{p,q,r}$ with $n = p+q+r = 4$ leaves orthogonal vectors orthogonal, though it does not preserve the inner product. This is due to the fact that for $n=4$, any vector anti-commutes with $\expval{S}_4$, and thus for any vector $u \in \GR[(1)]{p,q,r}$, we have $u S = \breve{S} u$, while $S X = X S$. As a consequence, the inner product between two vectors $u, v \in \GR[(1)]{p,q,r}$ is only scaled by the scalar $\Vert S^2 \Vert_\mathbb{S}$ under the transformation $y' \to X y \widetilde{X}$:
    \begin{align}
        u' \cdot v' &= \tfrac{1}{2} \pqty{u'v' + v' u'} = \tfrac{1}{2} \pqty{ X u \widetilde{X} X v \widetilde{X} + X v \widetilde{X} X u \widetilde{X}} \notag \\
        &= \tfrac{1}{2} \pqty{ X u S^2 v \widetilde{X} + X v S^2 u \widetilde{X}} = \tfrac{1}{2} \pqty{ X \breve{S}^2 u v \widetilde{X} + X \breve{S}^2 v u \widetilde{X}} \notag \\
        &= X \breve{S}^2 (u \cdot v) \widetilde{X} = S^2 \breve{S}^2 (u \cdot v) = \norm{S^2}_\mathbb{S} (u \cdot v)
    \end{align}
In particular, this means that orthogonal vectors remain orthogonal, since they satisfy $u \cdot v = 0$.
When $n > 4$ this construction fails, and we can no longer categorically state that all $X \in \GR[+]{p,q,r}$ respect orthogonality.
\end{remark}

\begin{example}[Bivector Generators]\label{ex:bivectors}
The bivector generators of rotations, translations and boosts in the geometric algebras \GR{2}, \GR{1,1}, and \GR{1,0,1} form an interesting example of normalization to consider, because they are elements of the Lie algebra, but only in the rotation case is the generator also an element of the group. Explicitly, when $\e{12}$ is a bivector satisfying $\e{12}^2 = -1$, it is the generator of a rotation, and thus $\exp(\tfrac{1}{2} \pi \e{12}) = \e{12}$. Hence, it is both an element of the Lie algebra and the Lie group, and thus we find for $X = \theta \e{12}$ that it normalizes to $R = \sign(\theta) \e{12}$ as expected:
    \begin{equation}
        R = \pqty{\theta^2}^{-1/2} \theta \e{12} = \frac{\theta}{\abs{\theta}} \e{12} = \sign(\theta) \e{12}.
    \end{equation}
However, when $\e{12}$ is the generator of a boost and thus $\e{12}^2 = 1$, it is not an element of the group, since there is no solution to $\exp(\theta \e{12}) \propto \e{12}$ for $\theta \in \mathbb{R}$. However, if we allow $\theta \in \mathbb{C}$, then clearly there is the solution $\theta = \tfrac 1 2 \pi i$. Indeed, from the normalization procedure, we find for $X = \theta \e{12}$ that $X \widetilde X = - \theta^2$ and thus
    \begin{equation}
        R = \pqty{- \theta^2}^{-1/2} \theta \e{12} = \frac{\theta}{\abs{\theta}} i \e{12} = \sign(\theta) i \e{12}.
    \end{equation}
This is a Lie group element since it satisfies the rotor condition $R \widetilde{R} = 1$ and we recognize it as again a rotation. Thus, the nearest group element to the boost generator $\e{12}$ is the rotation $i \e{12}$.
However, the action of $i \e{12}$ on an element $V$ of the algebra is identical to that of $\e{12}$ since the $i$'s cancel:
    \[ (i \e{12})[V] = i \e{12} V \e{12}^{-1} i^{-1} = \e{12} V \e{12}^{-1} = \e{12} [V]. \]
Hence, as transformations $i\e{12}$ and $\e{12}$ act identically, though they are different elements. 
If the goal is only to use $R$ as a transformation, one can get away with calculating $\vert X \widetilde X \vert^{-1/2}$ instead of $(X \widetilde X)^{-1/2}$, which is what is commonly done in GA texts and is also the approach taken in \cref{sec:numerical}.

This leaves only the generator of a translation, which is a bivector satisfying $\e{12}^2 = 0$, and thus for $X = \theta \e{12}$ we find $X \widetilde X = 0$. Such an element can therefore not be normalized, which can be interpreted to mean that no unambiguous nearest rotor to $X$ exists.
\end{example}

\section{Exponential Map}\label{sec:exponential}
A rotor $R$ generates an entire family of continous transformations upon exponation by a scalar $t$, since
    \begin{equation}
        R^{t} = e^{t \Log(R)}.
    \end{equation}
From this point of view the square root is merely the $t=\tfrac{1}{2}$ case.
To allow for arbitrary values of $t$, we need to have the exponential and logarithmic maps.

The exponential function maps elements of the Lie algebra to elements of the Lie group. Elements of the Lie algebra are always bivectors, a fact that for $n < 6$ follows directly from the normalization condition on $2k$-reflections. Consider a $2k$-reflection $R(t)$ satisfying $R(t) \widetilde{R}(t) = 1$. The derivative of this condition with respect to $t$ yields
    \begin{equation}
        \dot{R} \widetilde{R} + R \dot{\widetilde{R}} = 0,
    \end{equation}
and thus $\dot{R} \widetilde{R} = - R \dot{\widetilde{R}}$. 
However, $\widetilde{\dot{R} \widetilde{R}} = R \dot{\widetilde{R}}$, and thus we find that $\dot{R} \widetilde{R} = - \widetilde{\dot{R} \widetilde{R}}$, which in the even subalgebra for $n < 6$ is only satisfied for bivectors. (For $n \geq 6$ see e.g. \cite{LGasSG,roelfs2021graded}.)
Defining the bivector $\B := \dot{R} \widetilde{R}$, the differential equation can now be integrated to yield
    \begin{equation}
        R(t) = e^{t\B}.
    \end{equation}
Thus, the exponential function maps bivectors into $2k$-reflections.
In order to compute $R$ for a given bivector $\B$, the invariant decomposition can be used to decompose $\B$ into at most $k = \floor{n/2}$ commuting simple bivectors \cite{roelfs2021graded}.
For $n < 6$, any bivector can therefore be split into at most two simple bivectors, given by
    \begin{align}\label{eq:invariant_decomposition}
        \bl_\pm &= \frac{\B \cdot \B  + \B \wedge \B  \pm  \sqrt{\pqty{\B \cdot \B}^2 - \pqty{\B  \wedge \B}^2}}{2\B} \\
        &= \frac{\B^2 \pm \norm{\B^2}_\mathbb{S}}{2 \B} = \underbrace{\tfrac{1}{2} \pqty{1 \pm \frac{\breve{\B^2}}{\norm{\B^2}_\mathbb{S}}}}_{P_\pm}\B. \label{eq:invariant_decomposition2}
    \end{align}
such that $\B = \bl_+ + \bl_-$, and $\bl_+ \bl_- = \bl_+ \wedge \bl_-$ and thus $\bl_+ \cdot \bl_- = \bl_+ \times \bl_- = 0$,\footnote{$\bl_+ \times \bl_-$ is the commutator product: $\bl_+ \times \bl_- = \tfrac{1}{2} \pqty{\bl_+ \bl_- - \bl_- \bl_+}$.} with squares 
    \[ \lambda_\pm := \bl_\pm^2 = \tfrac{1}{2} \expval{\B^2} \pm \tfrac{1}{2} \norm{\B^2}_\mathbb{S}. \]
For the simplifications from \cref{eq:invariant_decomposition} to \cref{eq:invariant_decomposition2} we have used that $B^2$ is a Study number and $B^{-1} = B /B^2 = B \, \breve{B^2} / \Vert \B^2 \Vert_\mathbb{S}^2$.
The Study numbers 
\begin{equation}\label{eq:projectors}
    P_\pm(\B) = \tfrac{1}{2} \pqty{1 \pm \frac{\breve{\B^2}}{\norm{\B^2}_\mathbb{S}}}
\end{equation}
project $\B$ onto $\bl_\pm$.
Complex solutions can occur when
    \begin{align} \label{eq:imaginary_when}
        \Vert \B^2 \Vert_\mathbb{S}^2 &= \pqty{\B \cdot \B}^2 - \pqty{\B \wedge \B}^2 < 0 .
    \end{align}
Evidently this can only happen when $\pqty{\B \wedge \B}^2 > \pqty{\B \cdot \B}^2$, and because $\pqty{\B \cdot \B}^2 \geq 0$ is a non-negative real scalar, only spaces with $\pqty{\B \wedge \B}^2 > 0$ require further scrutiny.
Thus, we can immediately conclude that hyperbolic PGA (\GR{3,1}) and Eulidean PGA (\GR{3,0,1}) do not have complex solutions, since their quadvectors square to $0$ and $-1$ respectively. 
Furthermore, in elliptic PGA (\GR{4}) all the basis bivectors square to $-1$, and thus any bivector $\B$ can be brought to the form $\B' = a \e{12} + b \e{34}$ via an orthogonal transformation, where $a, b \in \mathbb{R}$ since $\B$ is a real bivector. This yields $\Vert \B'^2 \Vert_\mathbb{S}^2 = a^4 + b^4 - 2 a^2 b^2$, which is a non-negative function, and thus no imaginary solutions can appear in this space either.
Lastly, because any bivector in 3DCGA (\GR{4,1}) is isomorphic to either elliptic, hyperbolic or Euclidean PGA, it is also free from complex solutions.
The only space known to suffer from complex solutions is \GR{2,2}, see \cite[Example 6.4]{roelfs2021graded} for more details.

In order to exponentiate a bivector $\B$, the first step is to perform the invariant decomposition \cref{eq:invariant_decomposition} of $\B$ into the commuting simple bivectors $\bl_\pm$, after which the closed form exponential function follows straightforwardly from Euler's formula \cref{eq:euler}:
\begin{align}\label{eq:exponential}
    R &= e^\B = e^{\bl_+} e^{\bl_-} = \bqty{\co(\bl_+) + \si(\bl_+)} \bqty{\co(\bl_-) + \si(\bl_-)} = \co(\B) + \si(\B).
\end{align}

\section{Logarithmic Map}\label{sec:logarithm}
The logarithm is used to convert from a rotor $R \in \Spin{p,q,r}$ to a bivector $\B$ in the Lie algebra $\spin{p,q,r}$. To find it, we simply work backwards starting from \cref{eq:exponential} and \cref{eq:generalized_cos,eq:generalized_sin}. 
The invariant factorization, factors a $2k$-reflection $R \in \Spin{p,q,r}$ into $k$ commuting bireflections $R_i \in \Spin{p,q,r}$ \cite{roelfs2021graded}, and thus in $n < 6$ we find
    \begin{equation}\label{eq:invariant_factors}
        \begin{array}{r @{{}={}} c @{{}={}} c @{{}+{}} c}
            R & e^\B
            & \co(\B) & \si(\B) \\
            & e^{\bl_+}e^{\bl_-} & \overbrace{\co(\bl_+)\co(\bl_-) + \si(\bl_+)\si(\bl_-)} & \overbrace{\si(\bl_+)\co(\bl_-) + \co(\bl_+)\si(\bl_-)},
        \end{array}
    \end{equation}
where $\B = \bl_+ + \bl_-$ and $\bl_+ \bl_- = \bl_- \bl_+$. Because $e^{\bl_+}$ and $e^{\bl_-}$ commute, the principal logarithm $\B$ of $R$ can be defined as
    \begin{equation}\label{eq:logarithm}
	    \B := \Ln R = \Ln R_+ + \Ln R_- = \bl_+ + \bl_-,
\end{equation}
where in the case of rotations we assume $\theta_\pm := \sqrt{-\bl_\pm^2} \leq \pi$.
The logarithm $\B = \Ln R$ is related to the bivector part of the rotor $\expval{R}_2 = \si(\B)$ via a multiplicative factor $S$, such that
    \[ \B = S \si(\B). \]
This multiplicative factor $S$ is a self-reverse Study number, because \cref{eq:generalized_sin} dictates $\B \times \si(\B) = 0$, and thus
\begin{align}
    S = \B \si(\B)^{-1} = \B \cdot \si(\B)^{-1} + \cancel{\B \times \si(\B)^{-1}} + \B \wedge \si(\B)^{-1} 
\end{align}
has only scalar and quadvector parts.
Thus, if we can find the Study number $S = \alpha + \beta I$, the logarithm is given by $\B = S \si(\B)$. 
Via the invariant decomposition \cref{eq:invariant_factors}, $\B = S \si(\B)$ decomposes into two independent equations:
    \begin{align}
        \bl_+ &= \alpha \sico{\bl_+}{\bl_-} + \beta I \sico{\bl_-}{\bl_+} \\
        \bl_- &= \alpha \sico{\bl_-}{\bl_+} + \beta I \sico{\bl_+}{\bl_-}.
    \end{align}
These can be solved to find $\alpha$ and $\beta I$:
    \begin{alignat}{3}
        \alpha &= \B \cdot \si(\B)^{-1} &&= \frac{\bl_+ \si(\bl_+) \co(\bl_-) - \bl_- \si(\bl_-) \co(\bl_+)}{\norm{\si^2(\B)}_\mathbb{S}}, \\
        \beta I &= \B \wedge \si(\B)^{-1} &&= \frac{\bl_- \si(\bl_+) \co(\bl_-) - \bl_+ \si(\bl_-) \co(\bl_+)}{\norm{\si^2(\B)}_\mathbb{S}},
    \end{alignat}
where $\norm{\si^2(\B)}_\mathbb{S} = \sqrt{(\expval{R}_2 \cdot \expval{R}_2)^2 - 4 \expval{R}^2 \expval{R}_4^2}$ by using \cref{eq:norm_S}. 
In terms of the scalars $\theta_\pm := \sqrt{- \bl_\pm^2}$ this reduces to
    \begin{alignat*}{3}
        \alpha &= \frac{1}{\norm{\si^2(\B)}_\mathbb{S}} \bqty{\sin(\theta_-) \cos(\theta_+) \theta_- - \sin(\theta_+) \cos(\theta_-) \theta_+ }, \\
        \beta I &= \frac{1}{\norm{\si^2(\B)}_\mathbb{S}} \bqty{\sinc(\theta_+) \cos(\theta_-) - \sinc(\theta_-) \cos(\theta_+)} \bl_+ \bl_-.
    \end{alignat*}
Thus, for a given rotor $R$, the values of $\alpha$ and $\beta I$ can be calculated directly if we compute the values of $\theta_\pm$.
In order to extract $\theta_\pm$ we define the scalars $\sigma_\pm = \sqrt{\sico[2]{\bl_\pm}{\bl_\mp}}$.
In the algebras \GR{3,0,1}, \GR{3,1}, \GR{4} and \GR{4,1}, the simple bivector $\bl_-$ always corresponds to a rotation, and thus $\theta_-$ is given by
    \[ \theta_- = \arctantwo(\sigma_-, \expval{R}), \]
where $\arctantwo$ is the two-parameter tangent function.
With $\theta_-$ known, $\theta_+$ can be found from e.g. $\langle R \rangle$ or $\sigma_+$. Which strategy we choose to find $\sigma_+$ depends on the algebra, and can be seen in the numerical implementation. Note that for boosts, $\theta_+$ will be imaginary, mapping the trigonometric functions to their hyperbolic counterparts. Therefore $\alpha$ and $\beta I$ are real. In our numerical solutions below, this case is handled without the need for imaginary coefficients.

\section{Numerical Implementations}\label{sec:numerical}
We prepared efficient numerical implementations of normalization, exponential maps, and logarithmic maps in various 1-up and 2-up 3D geometries, which have been made available online \cite{sourcecode}.
These formulas are all symbolically optimized on the coefficient level, and therefore achieve optimal performance and accuracy. 
In this section we present only the normalization code, grouped per algebra. For 3DPGA the logarithmic and exponential maps are very short and also included.

\subsection{3D Hyperbolic PGA / Spacetime Algebra}

For $\mathbb R_{3,1}$, used as STA and 3D hyperbolic PGA, the normalisation can be further simplified. Starting from $R = \alpha X + \beta I X$, we define $X \widetilde X = s + tI$, and, using the formulas in the previous sections, can solve for $\alpha, \beta$ in terms of $s, t$ as

\[
\alpha = n^2m, \quad \beta = -tm 
\]
\[
n = \sqrt{\sqrt{s^2 + t^2} + s},\quad m= \cfrac {\sqrt{2}n}{n^4 + t^2}
\]
\noindent
Picking a basis and working things out at a coefficient level results in the following efficient implementation:
\begin{longlisting}
    \begin{minted}{javascript}
// STA/Hyperbolic PGA R3,1. e1*e1 = e2*e2 = e3*e3 = 1, e4*e4 = -1
// Normalize an even element X on the basis [1,e12,e13,e14,e23,e24,e34,e1234]
function Normalize(X) {
  var S = X[0]*X[0]+X[1]*X[1]+X[2]*X[2]-X[3]*X[3]+X[4]*X[4]-X[5]*X[5]-X[6]*X[6]-X[7]*X[7];
  var T = 2*(X[0]*X[7]-X[1]*X[6]+X[2]*X[5]-X[3]*X[4]);
  var N = ((S*S+T*T)**0.5+S)**0.5, N2 = N*N;
  var M = 2**.5*N/(N2*N2+T*T);
  var A = N2*M, B = -T*M;
  return rotor(A*X[0]-B*X[7], A*X[1]+B*X[6], A*X[2]-B*X[5], A*X[3]-B*X[4],
               A*X[4]+B*X[3], A*X[5]+B*X[2], A*X[6]-B*X[1], A*X[7]+B*X[0] );
}

// Square root of a rotor R;
function sqrt (R) { return Normalize(1 + R); }
    \end{minted}
\caption{Normalisation and square root of an element of the even subalgebra $R \in \GR[+]{3,1}$}
\label{lst:normalize31}
\end{longlisting}

\subsection{3D Euclidean PGA}

For $\GR{3,0,1}$, modeling the Euclidean group and geometry, the degenerate metric
provides even further opportunities for simplification. Starting again from $R = \alpha X + \beta I X$, and defining $X \widetilde X = s + tI$, we find directly
\[
\alpha = \cfrac {1} {\sqrt{s}},\quad \beta = \cfrac {t} {2\sqrt{s}^3},
\]
reducing the entire normalisation procedure to just 23 multiplications, 10 additions, 1 square root and 1 division. 
\begin{longlisting}
    \begin{minted}{javascript}
// 3D PGA. e1*e1 = e2*e2 = e3*e3 = 1, e0*e0 = 0
// Normalize an even element X on the basis [1,e01,e02,e03,e12,e31,e23,e0123]
function Normalize(X) {
  var A = 1/(X[0]*X[0] + X[4]*X[4] + X[5]*X[5] + X[6]*X[6])**0.5;
  var B = (X[7]*X[0] - (X[1]*X[6] + X[2]*X[5] + X[3]*X[4]))*A*A*A;
  return rotor( A*X[0], A*X[1]+B*X[6], A*X[2]+B*X[5], A*X[3]+B*X[4],
                A*X[4], A*X[5], A*X[6], A*X[7]-B*X[0] );
}

// Square root of a rotor R;
function sqrt(R) { return Normalize(1 + R); }

// Logarithm of a rotor R (14 mul, 5 add, 1 div, 1 acos, 1 sqrt)
function log(R) {
  if (R[0]==1) return bivector(R[1],R[2],R[3],0,0,0);
  var a = 1/(1 - R[0]*R[0]), b = acos(R[0])*sqrt(a), c = a*R[7]*(1 - R[0]*b);  
  return bivector( c*R[6] + b*R[1], c*R[5] + b*R[2], c*R[4] + b*R[3], b*R[4], b*R[5], b*R[6] );    
}

// Exponential of a bivector B (17 mul, 8 add, 2 div, 1 sincos, 1 sqrt)
function exp(B) {
  var l = (B[3]*B[3] + B[4]*B[4] + B[5]*B[5]);
  if (l==0) return rotor(1, B[0], B[1], B[2], 0, 0, 0, 0);
  var m = (B[0]*B[5] + B[1]*B[4] + B[2]*B[3]), a = sqrt(l), c = cos(a), s = sin(a)/a, t = m/l*(c-s);
  return rotor(c, s*B[0] + t*B[5], s*B[1] + t*B[4], s*B[2] + t*B[3], s*B[3], s*B[4], s*B[5], m*s);
}
     \end{minted}
\caption{Normalisation, Square root, Logarithmic and Exponential maps for $R, B \in \GR[+]{3,0,1}$.}
\label{lst:logexp301}
\end{longlisting}
\subsection{3D Elliptic PGA}
For $\mathbb R_{4}$, 3D elliptic PGA, the situation follows that of STA. Starting from $R = \alpha X + \beta I X$, we define $X\widetilde X = s + tI$, and, using the formulas in the previous sections, can solve for $\alpha, \beta$ in terms of $s, t$ as
\[
\alpha = n^2m, \quad \beta = -tm 
\]
\[
n = \sqrt{\sqrt{s^2 - t^2} + s},\quad m= \cfrac {\sqrt{2}n}{n^4 - t^2}
\]
\noindent
Picking a basis and working things out at a coefficient level results in:

\begin{longlisting}
    \begin{minted}{javascript}
// Elliptic/Spherical PGA. e1*e1 = e2*e2 = e3*e3 = e4*e4 = 1
// Normalize an even element X on the basis [1,e12,e13,e14,e23,e24,e34,e1234]
function Normalize(X) {
  var S = X[0]*X[0]+X[1]*X[1]+X[2]*X[2]+X[3]*X[3]+X[4]*X[4]+X[5]*X[5]+X[6]*X[6]+X[7]*X[7];
  var T = 2*(X[0]*X[7]-X[1]*X[6]+X[2]*X[5]-X[3]*X[4]);
  var N = ((S*S-T*T)**0.5+S)**0.5, N2 = N*N;
  var M = 2**.5*N/(N2*N2-T*T);
  var A = N2*M, B = -T*M;
  return rotor( A*X[0]+B*X[7], A*X[1]-B*X[6], A*X[2]+B*X[5], A*X[3]-B*X[4],
                A*X[4]-B*X[3], A*X[5]+B*X[2], A*X[6]-B*X[1], A*X[7]+B*X[0] );
}

// Square root of a rotor R;
function sqrt(R) { return Normalize(1 + R); }
     \end{minted}
\caption{Normalisation and square root of an element of the even subalgebra $R \in \GR[+]{4}$}
\label{lst:normalize4}
\end{longlisting}
\subsection{3D Conformal GA}
For the 5 dimensional $R_{4,1}$, modelling the 3D conformal group and geometry, the situation remains unchanged, although the resulting coordinate expressions are slightly more involved. Following the procedure from STA, we now get $X\widetilde X = s + t_i\mathbf e_{i}I$ where the $t_i$ are the 5 coefficients of the grade 4 part and $I$ is the pseudoscalar. We find
\[
\alpha = n^2m, \quad \beta_i = -t_im 
\]
\[
n = \sqrt{\sqrt{s^2 - t_1^2 + t_2^2 + t_3^2 + t_4^2 + t_5^2} + s},\quad m= \cfrac {\sqrt{2}n}{n^4  - t_1^2 + t_2^2 + t_3^2 + t_4^2 + t_5^2}
\]
\begin{longlisting}
    \begin{minted}{javascript}
// CGA R4,1. e1*e1 = e2*e2 = e3*e3 = e4*4 = 1, e5*e5 = -1
// Normalize an even element X = [1,e12,e13,e14,e15,e23,e24,e25,e34,e35,e45,e1234,e1235,e1245,e1345,e2345]
function Normalize(X) {
  var S  = X[0]*X[0]-X[10]*X[10]+X[11]*X[11]-X[12]*X[12]-X[13]*X[13]-X[14]*X[14]-X[15]*X[15]+X[1]*X[1]
          +X[2]*X[2]+X[3]*X[3]-X[4]*X[4]+X[5]*X[5]+X[6]*X[6]-X[7]*X[7]+X[8]*X[8]-X[9]*X[9];
  var T1 = 2*(X[0]*X[11]-X[10]*X[12]+X[13]*X[9]-X[14]*X[7]+X[15]*X[4]-X[1]*X[8]+X[2]*X[6]-X[3]*X[5]);
  var T2 = 2*(X[0]*X[12]-X[10]*X[11]+X[13]*X[8]-X[14]*X[6]+X[15]*X[3]-X[1]*X[9]+X[2]*X[7]-X[4]*X[5]);
  var T3 = 2*(X[0]*X[13]-X[10]*X[1]+X[11]*X[9]-X[12]*X[8]+X[14]*X[5]-X[15]*X[2]+X[3]*X[7]-X[4]*X[6]);
  var T4 = 2*(X[0]*X[14]-X[10]*X[2]-X[11]*X[7]+X[12]*X[6]-X[13]*X[5]+X[15]*X[1]+X[3]*X[9]-X[4]*X[8]);
  var T5 = 2*(X[0]*X[15]-X[10]*X[5]+X[11]*X[4]-X[12]*X[3]+X[13]*X[2]-X[14]*X[1]+X[6]*X[9]-X[7]*X[8]);
  var TT = -T1*T1+T2*T2+T3*T3+T4*T4+T5*T5;
  var N = ((S*S+TT)**0.5+S)**0.5, N2 = N*N;
  var M = 2**0.5*N/(N2*N2+TT);
  var A = N2*M, [B1,B2,B3,B4,B5] = [-T1*M,-T2*M,-T3*M,-T4*M,-T5*M];
  return rotor(A*X[0]  + B1*X[11] - B2*X[12] - B3*X[13] - B4*X[14] - B5*X[15],
               A*X[1]  - B1*X[8]  + B2*X[9]  + B3*X[10] - B4*X[15] + B5*X[14],
               A*X[2]  + B1*X[6]  - B2*X[7]  + B3*X[15] + B4*X[10] - B5*X[13],
               A*X[3]  - B1*X[5]  - B2*X[15] - B3*X[7]  - B4*X[9]  + B5*X[12],
               A*X[4]  - B1*X[15] - B2*X[5]  - B3*X[6]  - B4*X[8]  + B5*X[11],
               A*X[5]  - B1*X[3]  + B2*X[4]  - B3*X[14] + B4*X[13] + B5*X[10],
               A*X[6]  + B1*X[2]  + B2*X[14] + B3*X[4]  - B4*X[12] - B5*X[9],
               A*X[7]  + B1*X[14] + B2*X[2]  + B3*X[3]  - B4*X[11] - B5*X[8],
               A*X[8]  - B1*X[1]  - B2*X[13] + B3*X[12] + B4*X[4]  + B5*X[7],
               A*X[9]  - B1*X[13] - B2*X[1]  + B3*X[11] + B4*X[3]  + B5*X[6],
               A*X[10] + B1*X[12] - B2*X[11] - B3*X[1]  - B4*X[2]  - B5*X[5],
               A*X[11] + B1*X[0]  + B2*X[10] - B3*X[9]  + B4*X[7]  - B5*X[4],
               A*X[12] + B1*X[10] + B2*X[0]  - B3*X[8]  + B4*X[6]  - B5*X[3],
               A*X[13] - B1*X[9]  + B2*X[8]  + B3*X[0]  - B4*X[5]  + B5*X[2],
               A*X[14] + B1*X[7]  - B2*X[6]  + B3*X[5]  + B4*X[0]  - B5*X[1],
               A*X[15] - B1*X[4]  + B2*X[3]  - B3*X[2]  + B4*X[1]  + B5*X[0]);
}

// Square root of a rotor R;
function sqrt(R) { return Normalize(1 + R); }
     \end{minted}
\caption{Normalisation and square root of an element of the even subalgebra $R \in \GR[+]{4,1}$}
\label{lst:normalize41}
\end{longlisting}

\section{Conclusion}
We presented closed form solutions for renormalization, square roots and exponential and logarithmic maps for elements for all geometric algebras \GR[+]{p,q,r} with $n = p+q+r < 6$. 
Concrete efficient implementations for the most popular such algebras \GR[+]{4}, \GR[+]{3,1}, \GR[+]{3,0,1} and \GR[+]{4,1}, were presented in order to facilitate immediate implementation in the library of your choice.

Because of a formal analysis of Study numbers, the current work was able to present a signature agnostic formulation of renormalization, based on the principles previously laid out in \cite{roelfs2021graded}.
While the limit of $n = 6$ was not broken, the presented approach provides a great deal of insight which can help with interesting future research into $n = 6$ and beyond.

\begin{acks}
The authors would like to thank dr. ir. Leo Dorst for invaluable discussions about this research.
The research of M.~R. was supported by 
\grantsponsor{1}{KU Leuven}{} IF project \grantnum[]{1}{C14/16/067}.
\end{acks}

\bibliographystyle{ACM-Reference-Format}
\bibliography{biblio.bib}

\end{document}